\documentclass[12pt]{article}

\usepackage{newtxtext,newtxmath}
\usepackage{graphicx}
\usepackage[letterpaper,margin=1in]{geometry}
\usepackage{xcolor} 
\renewenvironment{abstract}
	{\quotation}
	{\endquotation}

\date{}

\makeatletter
\renewcommand{\fnum@figure}{\textbf{Figure \thefigure}}
\renewcommand{\fnum@table}{\textbf{Table \thetable}}
\makeatother

\usepackage{scicite}
\usepackage{url}

\def\scititle{
	Controlled topological dilution drives cooperative glassy dynamics in artificial spin ice
}
\title{\bfseries \boldmath \scititle}

\author{
    Davis~Crater$^{1\dagger}$,
    Ryan~Mueller$^{1\dagger}$,
    Sanjib~Thapa$^{1}$,
    Kevin~Hofhuis$^{2}$,\and
    Armin~Kleibert$^{2}$,
    Francesco~Caravelli$^{3}$,
	Alan~Farhan$^{1\ast}$\and
    \small$^{1}$Department of Physics \& Astronomy, Baylor University, Waco, TX, 76798 USA.\and
    \small$^{2}$Paul Scherrer Institut, Forschungstrasse 111, 5232 Villigen PSI, Switzerland.\and
	\small$^{3}$ Theoretical Division (T-4), \small Los Alamos National Laboratory, Los Alamos, NM, 87545, USA.\and
	\small$^\ast$Corresponding author. Email: alan\textunderscore farhan@baylor.edu\and
	\small$^\dagger$These authors contributed equally to this work.
}

\begin{document}
\maketitle

\begin{abstract} \bfseries \boldmath
It has long been known that disorder, perturbing the energy landscape of magnetic systems, can introduce glassy dynamics. However, the controlled role of increasing disorder in driving glass formation remains difficult to isolate in naturally occurring materials. Artificial spin ice offers a unique model platform in which geometry, interactions, and disorder can be engineered at the nanoscale. Here, we investigate the impact of controlled disorder introduced through random decimation in artificial square spin ice. By systematically removing nanomagnets from random sites, we modify the vertex topology and progressively increase frustration in the spin network. Synchrotron-based photoemission electron microscopy reveals that decimation enhances the population of higher energy vertices and increases the configurational entropy of the system. Time-resolved temperature-dependent imaging further shows the emergence of slow cooperative dynamics at higher decimation, characterized by aging, a finite Edwards--Anderson order parameter, and enhanced dynamical heterogeneity quantified by the four-point susceptibility. The relaxation dynamics transition from thermally activated behavior at low decimation to Vogel--Fulcher--type freezing at higher decimation. These results demonstrate that random decimation drives artificial spin ice from long-range order to a glass-like magnetic state, establishing artificial spin systems as a tunable platform for studying glassy dynamics in frustrated matter.
\end{abstract}

\noindent
Disordered systems often display remarkably slow and complex dynamics that resemble the behavior of glasses~\cite{Binder1986,Debenedetti2001}. The degrees of freedom in such systems explore a rugged energy landscape containing many metastable states, leading to phenomena such as aging, dynamical heterogeneity, and non-Arrhenius-type relaxation~\cite{Binder1986,Debenedetti2001,Berthier2011,Ediger2000}. Glassy dynamics appear in a wide range of physical systems such as structural glasses, spin glasses, granular materials and colloidal systems~\cite{Ediger2000,Angell1995,Cavagna2009}. Despite decades of study, the microscopic mechanisms underlying glass formation remain difficult to access, because disorder, interactions, and lattice geometry are usually intertwined in naturally occurring materials. In the case of nanomagnetic systems like artificial spin ice, the relaxation to perturbations has already been investigated theoretically and numerically, arguing that in 2D \cite{Saccone20232} and 3D pyrochlore spin ice \cite{Halln2022}, fractal and anomalous noise can emerge. In this respect, magnetic systems provide a particularly rich playground to explore glass formation, because competing interactions in these systems can produce highly frustrated states of matter. Frustration occurs when interactions, subject to local constraints, are hindered from being simultaneously minimized. This results in extensive degeneracies of low-energy states and associated emergent phenomena~\cite{Harris2020}. Spin glasses are a special form of magnetic frustration emerging from disordered or randomized magnetic interactions. One limitation of naturally occurring spin glasses is that the disorder is intrinsic to the material itself, making controlled tuning difficult. This makes it challenging to disentangle the roles of disorder and geometry in producing glassy behavior.

Artificial spin ice provides a powerful platform to directly study the consequences of frustrated interactions in controlled settings~\cite{Skjaervo2020,Farhan2017,Farhan2019,Gilbert2014,Gilbert2015,Lao2018}. These systems consist of arrays of dipolar-coupled single-domain nanomagnets that are lithographically defined onto a variety of two-dimensional lattices. The shape of the nanomagnets can even be tuned to mimic Ising-, XY- or Potts-type spins~\cite{Farhan2019,Leo2018,Louis2018}. The main attraction of these artificial frustrated systems is that their magnetic configurations can be directly visualized using appropriate magnetic imaging techniques, providing direct real-space access to frustration-driven emergent phenomena, ranging from emergent magnetic monopoles~\cite{Farhan2019,Perrin2016,Ladak2010}, reduced- and elevated effective dimensionalities~\cite{Gilbert2015,Saccone2020} to dynamical glass transitions in disordered nanomagnetic networks~\cite{Saccone2022}.

A key advantage of artificial spin ice is that disorder can be introduced deliberately and systematically. In particular, random decimation---achieved by removing individual nanomagnets from the lattice---modifies the connectivity of the spin network and introduces vertices with reduced coordination. Although detailed studies on spin glasses and frustrated magnets have suggested that decimation can strongly influence the collective behavior of frustrated spin systems~\cite{Sen2015,Binder1986,Sala2014}, experimental investigations of how controlled decimation modifies both the thermodynamic landscape and the dynamical properties in frustrated systems, in general, and in artificial spin ice, in particular, have not yet been explored.

\begin{figure}
	\centering
	\includegraphics[width=0.95\columnwidth]{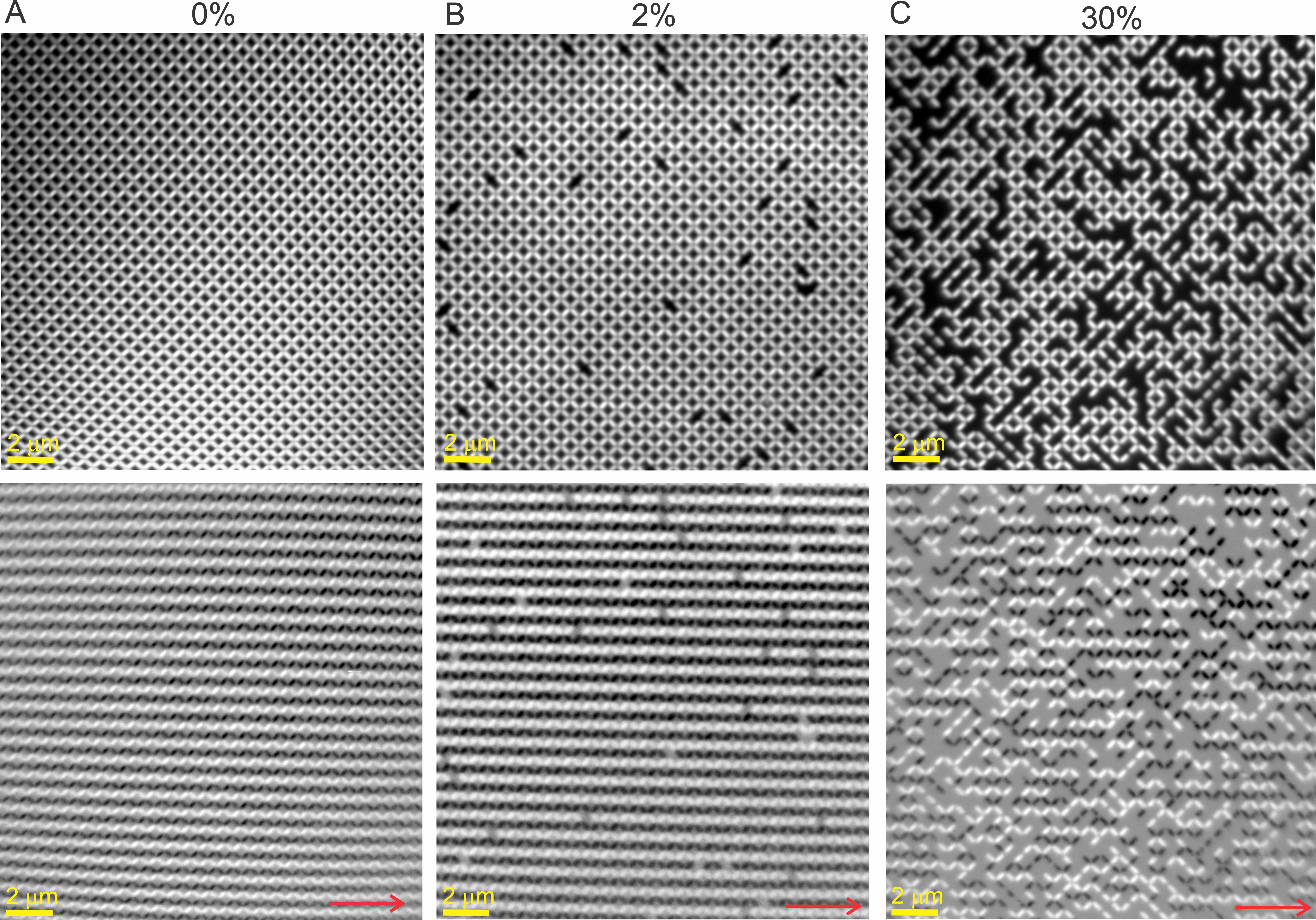}
	\caption{\label{fig:1} {\bf Random decimation of artificial spin ice}. Upper panel: single-polarization (left-circular polarization) PEEM imaging of artificial spin ice with increasing decimation starting from {\bf (A)} 0$\%$ through {\bf (B)} 2$\%$ to {\bf (C)} 30$\%$. Lower panel: corresponding XMCD images of low-energy states achieved after thermal annealing.}
\end{figure}

Here, we investigate how random decimation drives the emergence of glassy dynamics in artificial spin ice. This is achieved by placing Ising-type nanomagnets with lengths $L=400$~nm and widths $W=100$~nm onto a square lattice with a lattice parameter $a=550$~nm (see Methods). While the background square lattice and nanomagnet dimensions are maintained, we generate lattices in which nanomagnets are removed from random sites. The percentage of nanomagnet removal (decimation percentage) is varied to form eight decimation levels of randomly decimated artificial square ice, namely 0$\%$, 2$\%$, 5$\%$, 10$\%$, 15$\%$, 20$\%$, 25$\%$ and 30$\%$ (see examples in Fig.~\ref{fig:1}). Magnetic configurations and temperature-dependent fluctuations are visualized using synchrotron-based photoemission electron microscopy (PEEM), employing X-ray magnetic circular dichroism (XMCD) at the Fe L$_3$ edge~\cite{Stohr1993} (see Methods).

\section*{Results}
\section{Thermal annealing}
\begin{figure}
	\centering
	\includegraphics[width=0.85\columnwidth]{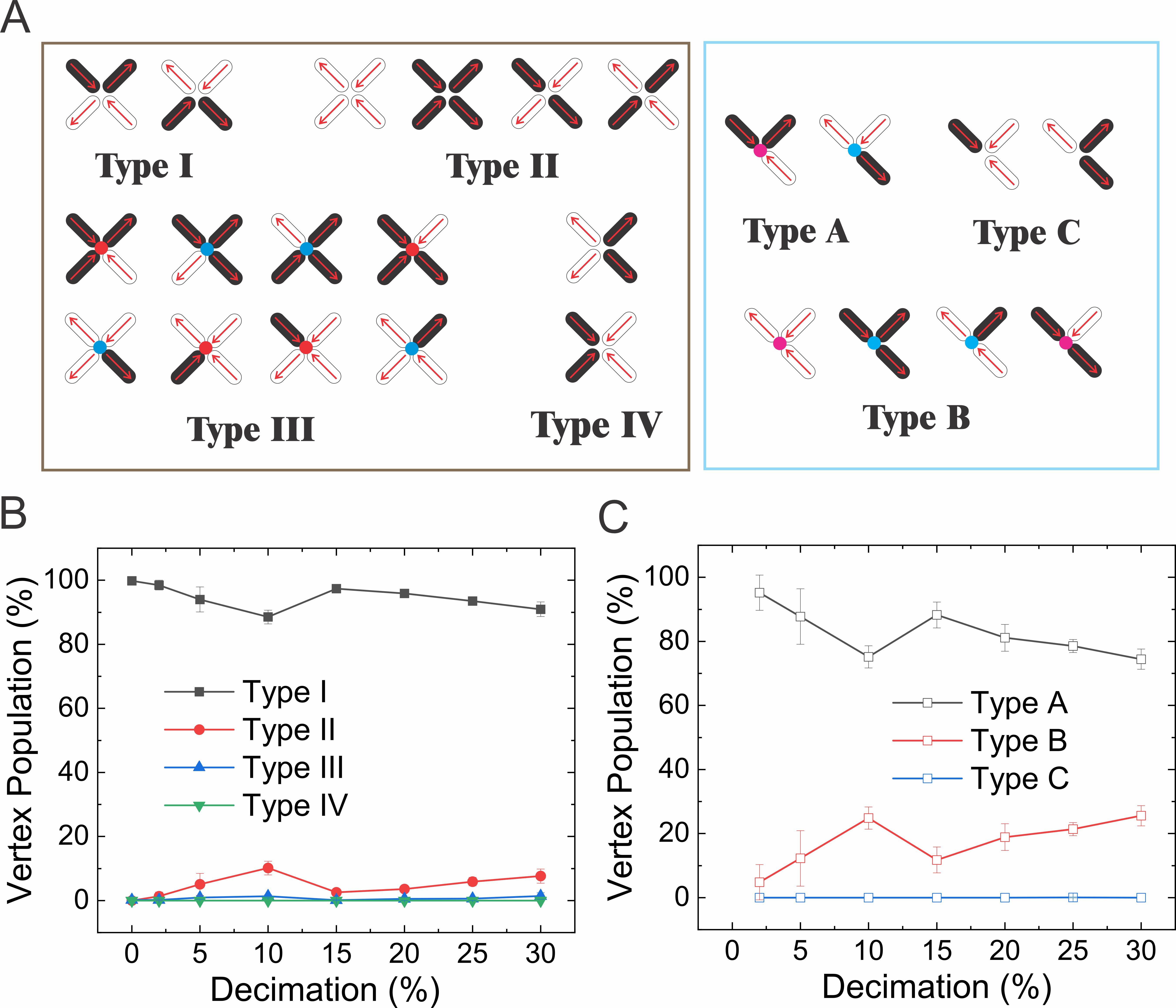}
	\caption{\label{fig:2} {\bf Vertex types and their decimation-dependent population}. {\bf (A)} Possible vertex types at four- and three-nanomagnet vertices listed with increasing dipolar energy. Using the dumbbell model~\cite{Moller2009}, Type III defects act as effective magnetic charges with either $+2q$ (red dots) or $-q$ (blue dots) residing at the vertex, while Type A and Type B vertices carry either $+q$ (magenta dots) or $-q$ (cyan blue dots) vertex charges. Type C and Type IV vertices are experimentally never observed. {\bf (B)} Vertex type population at four-nanomagnet vertices plotted as a function of increasing decimation percentage, achieved after thermal annealing. A clear dominance of ground state Type I vertices is observed.  {\bf (C)} Populations achieved at three-nanomagnet vertices. While Type A vertices are clearly dominating with values above 90$\%$ at lower decimation percentages, Type B vertices rise to around 25$\%$ in population at higher decimation percentages. The error bars in {\bf (B)} and {\bf (C)} represent standard deviations obtained from averaging over up to eight repetitions of the same annealing protocol.}
\end{figure}

As a first step, we apply a thermal annealing protocol to the fabricated structures~\cite{Farhan2019,Crater2024}, in which we keep the nanofabricated samples at room temperature inside a vacuum box for several weeks. Given a blocking temperature of around 200~K, this protocol ensures that the structures are properly thermalized and annealed before transfer to the PEEM instrument, where they are cooled to 150~K for imaging of the frozen low-energy states achieved after thermal annealing (see examples in Fig.~\ref{fig:1}). A clear indication of the effectiveness of this annealing protocol is the achievement of a 100$\%$ Type I ground state in the 0$\%$ decimated square ice (see Fig.~\ref{fig:1}A, lower panel). Analyzing the imaged configurations in terms of vertex types achieved at three- and four-nanomagnet vertices (see Fig.~\ref{fig:2}A), we extract the average vertex type populations and plot them as a function of decimation percentage (see Fig.~\ref{fig:2}B and Fig.~\ref{fig:2}C). Looking at these plots, we see a clear dominance of Type I vertices at the four-nanomagnet vertex sites, with Type II vertices remaining near 10$\%$ population throughout all decimation regimes (Fig.~\ref{fig:2}B). Type III defects are rarely observed. Three-nanomagnet vertices start to appear at decimation levels of 2$\%$ and above and show a clear dominance of Type A vertices at lower (sub-10$\%$) decimation percentages. Type B vertices increase toward 25$\%$ in population for decimation percentages of 10$\%$ and above, while Type C vertices are never observed (see Fig.~\ref{fig:2}C). A rise in Type B vertices is one of the typical features of vertex-frustrated artificial spin ice systems~\cite{Gilbert2014,Gilbert2015,Crater2024,Saccone2023,Morrison2013}. It indicates the emergence of modest vertex frustration with increasing random decimation. The non-monotonic evolution of the Type B population reflects a competition between local defect energetics and collective ordering. At intermediate dilution (~10$\%$), isolated vacancies favor Type B configurations to minimize local dipolar energy. Near 15$\%$, the emergence of larger correlated domains partially restores Type A-like local order, temporarily suppressing Type B vertices. At still higher dilution, interactions between neighboring defects and the increasingly frustrated charge network again stabilize Type B configurations, leading to their renewed increase, as can be seen in Figure~\ref{fig:2}C.

\begin{figure}
	\centering
	\includegraphics[width=0.9\columnwidth]{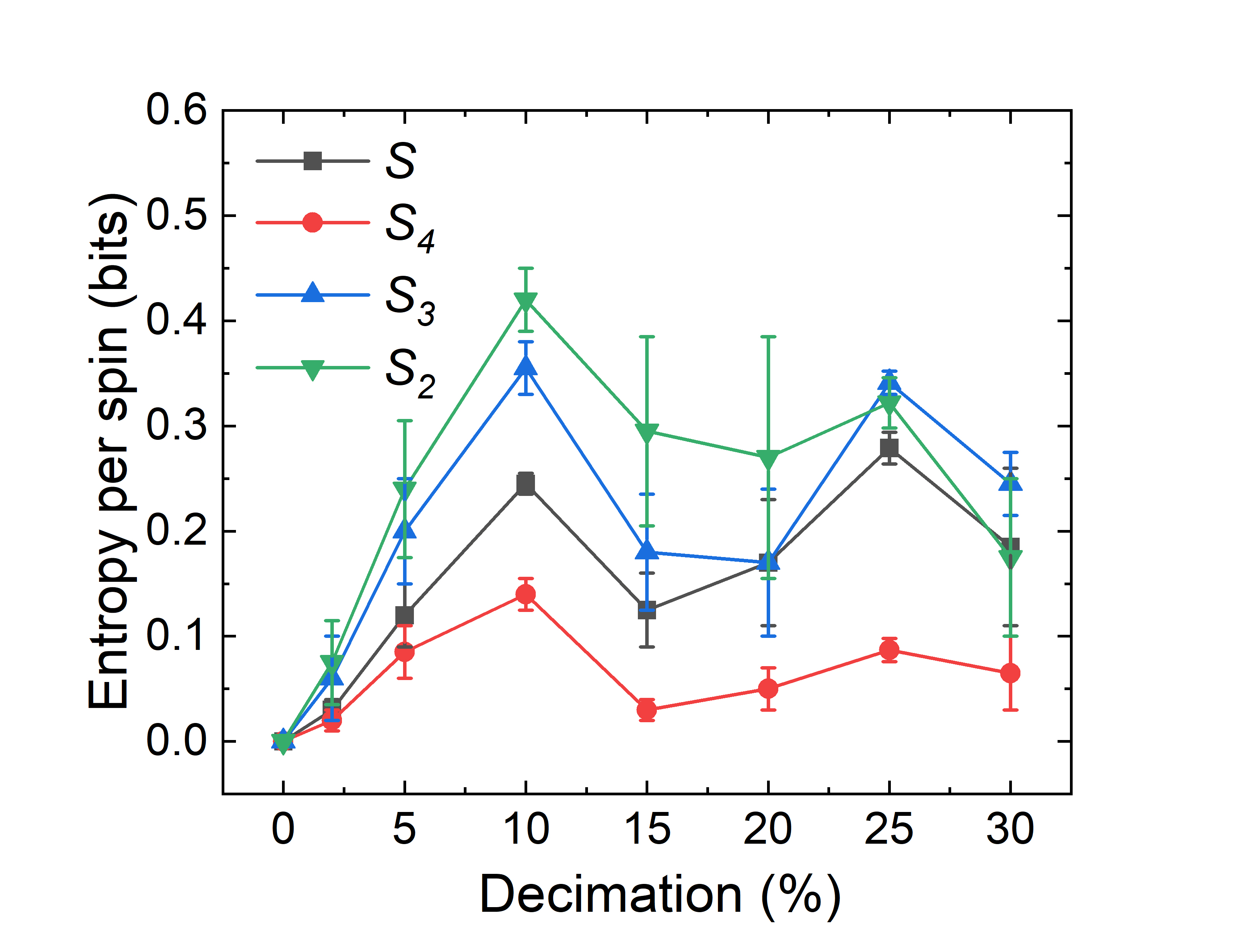}
	\caption{\label{fig:3} Entropies extracted as a function of decimation percentage.}
\end{figure}

To further confirm the rise in frustration with increasing decimation, we extract entropies~\cite{Lammert2010,Saccone2023} (see Methods) from moment configurations recorded as a function of decimation percentage (see Fig.~\ref{fig:3}). The observed trends confirm the conclusions drawn from the vertex-type populations: entropies rise with increasing decimation and reach near saturation at 10$\%$ decimation and above. These trends reflect a rise in disordered patterns, mostly at decimation sites, not involving four-nanomagnet vertices. In other words, four-nanomagnet vertices remain largely long-range ordered with a dominance of Type I vertices, while decimation sites give rise to frustration and disordered moment configurations. Furthermore, when employing the magnetic dumbbell model~\cite{Moller2009,Kagome2017}, where each magnetic moment $m$ of a nanomagnet with a length $L$ is replaced by a pair of magnetic charges $\pm q$, with $q= \mid m/L \mid$ residing at the ends of the nanomagnets, we observe a transition from highly charge-ordered patterns at low decimation to increasingly disordered vertex charge arrangements and a decay in charge-charge correlations (see supplementary materials and Figure S1).

\section{Thermal fluctuations and dynamic heterogeneity}
The consistent dominance of Type I vertices at four-nanomagnet vertices indicates that those sites are largely frozen once they fall into long-range-ordered Type I configurations, consistent with the confinement expected for the two-dimensional square-ice ground state~\cite{Moller2006,Alan2013,Budrikis2010}. Contrary to these frozen ground state regions, decimation sites feature an increase in high-energy Type B vertices and a general increase in disordered moment configurations, indicating that most dynamics occur at those sites. Taken together, this clearly points to dynamical heterogeneity and glassiness as decimation increases. 
To test this hypothesis, we performed temperature-dependent XMCD imaging of thermal fluctuations in artificial spin ice patterns with 15$\%$ and 30$\%$ decimation (see examples in movies S1 and S2).

\begin{figure}
	\centering
	\includegraphics[width=0.6\columnwidth]{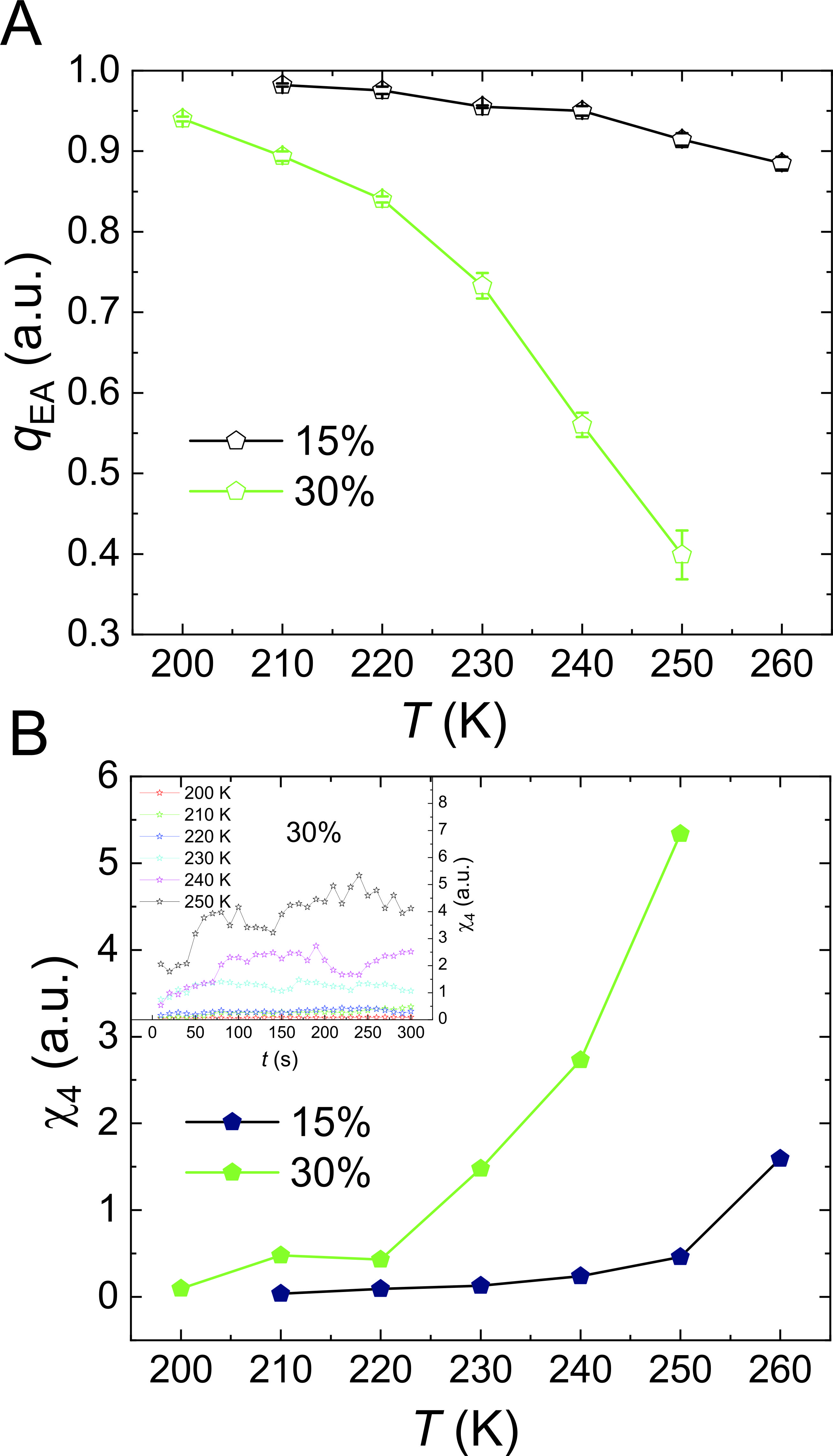}
	\caption{\label{fig:4} {\bf Glassiness and dynamical heterogeneity in decimated artificial spin ice}. {\bf (A)} Edwards-Anderson order parameter $q_{EA}$ extracted as a function of temperature for the 15$\%$ (black open pentagons and lines) and 30$\%$ (green open pentagons and lines) decimated artificial spin ice. {\bf (B)} Four-point susceptibility $\chi_{4}$ maxima derived from time-dependent $\chi_{4}$ curves (inset) within the same temperature range. Error bars represent standard deviations obtained from sequences of 80 XMCD images recorded at each temperature.}
\end{figure}

To quantitatively characterize the emergence of glassy dynamics suggested by the spatial heterogeneity observed at decimation sites, we evaluate the Edwards-Anderson order parameter $q_{EA}$ and four-point susceptibility $\chi_{4}$ (see Methods), both of which are widely used to identify slow dynamics and cooperative behavior in glassy systems~\cite{Edwards1975,Berthier2005}. $q_{EA}$ measures the persistence of a magnetic configuration over time and therefore quantifies the degree of freezing in the system. By contrast, the four-point susceptibility measures fluctuations in the two-time autocorrelation function and therefore probes the degree of dynamical heterogeneity. Large values of $\chi_{4}$ indicate that groups of spins undergo correlated rearrangements rather than fluctuating independently. 

Figure~\ref{fig:4}A shows the temperature dependence of the Edwards-Anderson order parameter for 15$\%$ and 30$\%$ decimated arrays. For the 15$\%$ system, $q_{EA}$ remains close to unity across the entire temperature range. This indicates that the majority of magnetic moments retain strong temporal correlations and remain effectively frozen on the experimental timescale. In contrast, the 30$\%$ decimated system features a pronounced decrease of $q_{EA}$ with increasing temperature, dropping significantly above 230~K. This behavior reflects a progressive loss of temporal correlations and indicates that spins in the highly-decimated spin ice system explore a larger portion of phase space. The reduced $q_{EA}$ in the 30$\%$-system signals the presence of enhanced fluctuations and a departure from a largely frozen state.

Further insight into the cooperative nature of these dynamics is gained by the four-point susceptibility $\chi_{4}$ (see Methods), shown in Figure~\ref{fig:4}B. While the 15$\%$ system displays only a modest increase in $\chi_{4}$ with temperature, the 30$\%$ system exhibits a strong and rapid growth of $\chi_{4}$, reaching significantly larger values across the entire temperature range. This indicates the emergence of spatially correlated fluctuations, i.e., clusters of spins undergoing cooperative rearrangements. Such behavior is a hallmark of dynamical heterogeneity and is commonly associated with glass-forming systems. Taken together, the contrasting behavior of $q_{EA}$ and $\chi_{4}$ for both decimation levels reveals a clear evolution of the dynamical state of the system. At moderate decimation (15$\%$), the dynamics remain largely governed by localized, thermally-activated processes, consistent with a weakly perturbed energy landscape. In contrast, at higher decimation (30$\%$), the system exhibits signatures of cooperative freezing, where extended regions of magnetic moments become dynamically correlated and relaxation proceeds through collective rearrangements.

\begin{figure}
	\centering
	\includegraphics[width=0.55\columnwidth]{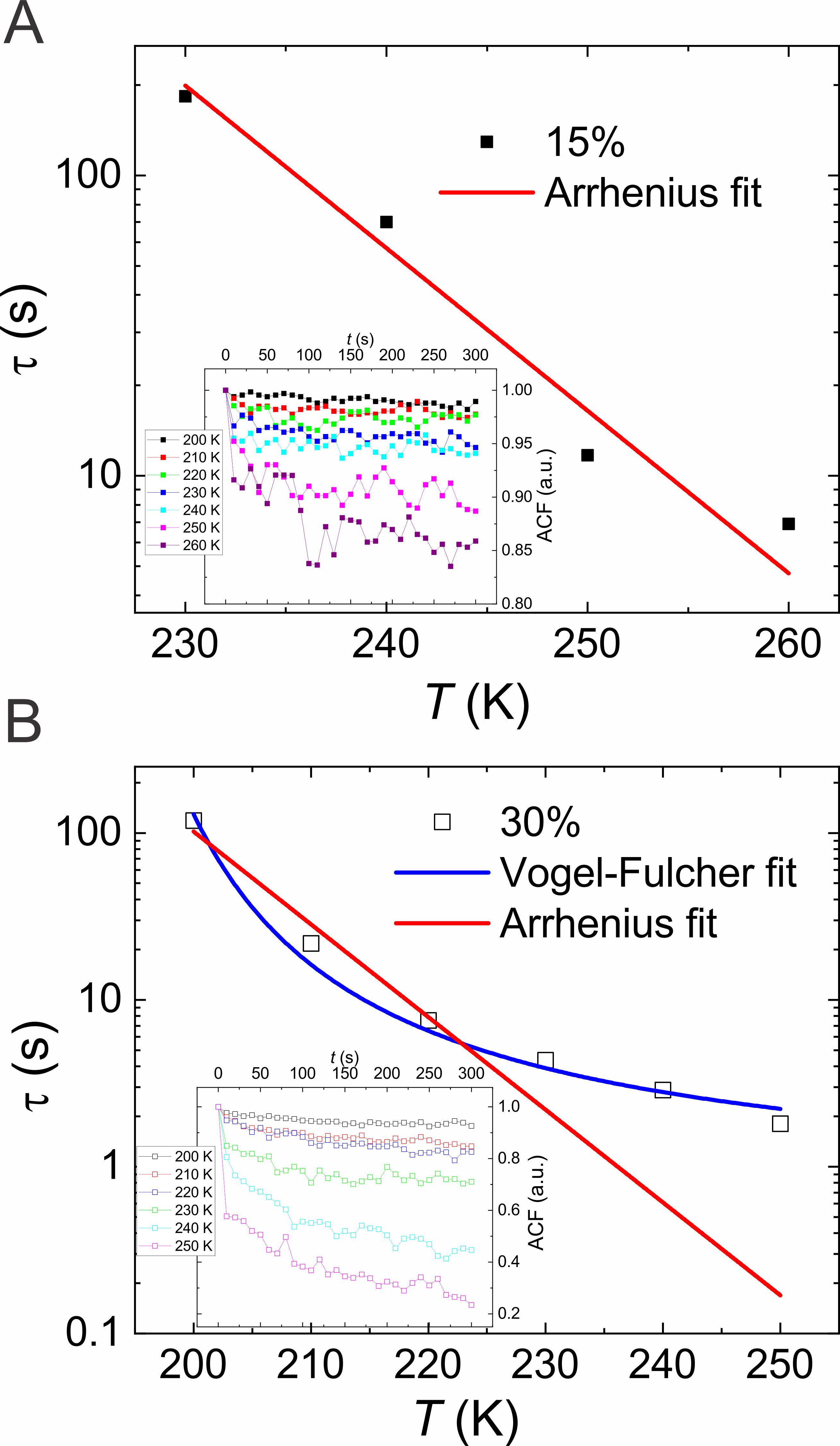}
	\caption{\label{fig:5} {\bf Arrhenius vs. Vogel-Fulcher relaxation}. {\bf (A)} Relaxation time extracted from the temporal evolution of the spin auto-correlation function (inset) and plotted as a function of temperature. Employing a $C(\tau)=0.95$ criterion, we excluded lower temperature curves as they do not decay below 0.95 within the experimental observation window, reflecting that the system remains effectively frozen. The red line is an Arrhenius fit, revealing a good match with experimentally extracted relaxation times {\bf (B)} Relaxation times extracted from the auto-correlation function decay (see inset) showing a good agreement with Vogel-Fulcher-type temperature dependence (blue curve), whereas the Arrhenius fit (red curve) does not match anymore.}
\end{figure}

To further quantify the nature of the relaxation dynamics, we extract characteristic relaxation times from temporal spin auto-correlation functions (see insets in Fig.~\ref{fig:5}) and examine their temperature dependence. Figure~\ref{fig:5} summarizes the resulting relaxation times for the 15$\%$ and 30$\%$ decimated systems, together with fits to Arrhenius and Vogel-Fulcher forms.

For the 15$\%$ decimated system (Fig.~\ref{fig:5}A), the relaxation time exhibits a temperature dependence that is well described by an Arrhenius form of activation $\tau =\tau_{0}exp(E/k_{B}T)$, with an activation energy $E=$ 0.43~eV and an attempt time $\tau_{0}=10^{-9}$~s. This behavior indicates that relaxation in the 15$\%$ decimated system proceeds predominantly through localized spin-flip events, governed by a well-defined energy barrier, consistent with the weak signatures of dynamical heterogeneity observed in Figure~\ref{fig:4}.

In contrast, the 30$\%$ decimated system (Fig.~\ref{fig:5}B) displays a pronounced deviation from Arrhenius behavior. While a simple activated description fails to capture the temperature dependence of the relaxation time, it fits well into a Vogel-Fulcher form, $\tau =\tau_{0}exp[A/(T-T_{0})]$, with a finite freezing temperature $T_{0}=$~184~K and activation scale $A=$~84~K. The emergence of Vogel-Fulcher behavior indicates that relaxation is no longer governed by a single energy barrier but instead reflects cooperative dynamics in a complex energy landscape with increasing barriers as the system approaches kinetic arrest. Previous work reported Vogel–Fulcher dynamics in a weakly interacting artificial square spin ice in which the lattice spacing was nearly four times larger than the nanomagnet length, strongly suppressing dipolar coupling. In that regime, the finite freezing temperature was interpreted as arising from weak residual interactions between nearly independent macrospins~\cite{Morley2017}. The present system is fundamentally different. Here, the nanomagnets remain strongly dipolar coupled and the interaction scale is unchanged across all decimation levels. Vogel–Fulcher behavior emerges only after random decimation introduces a sufficiently dense network of vacancies and reduced-coordination vertices. The accompanying rise in $\chi_{4}$ and decrease in $q_{EA}$ demonstrate that the slowing down is not simply due to weak interactions, but instead reflects cooperative many-body dynamics generated by defect-mediated frustration.

The crossover from Arrhenius to Vogel--Fulcher behavior with increasing decimation provides direct evidence for a qualitative change in the underlying relaxation mechanism. At moderate decimation, the system remains in a regime where dynamics are dominated by localized excitations. However, at higher decimation, the increasing density of defects and the resulting modification of the interaction network lead to the formation of correlated regions whose relaxation requires collective rearrangements. This transition is consistent with the strong enhancement of dynamical heterogeneity observed in Fig.~\ref{fig:4} and indicates the onset of glass-like freezing in the highly decimated arrays. Taken together, the combined analysis of relaxation times, dynamical heterogeneity, and temporal correlations establishes a coherent picture in which random decimation drives the system from a regime of independent, thermally activated dynamics to one characterized by cooperative, glass-like behavior. The ability to tune this crossover through controlled decimation highlights the role of connectivity and defect-induced interactions in shaping the energy landscape of frustrated magnetic systems.

\section*{Discussion}
The combined structural and dynamical measurements presented here establish a coherent picture in which random decimation fundamentally reshapes the energy landscape of artificial spin ice and drives the emergence of glassy dynamics. While the fully connected lattice remains dominated by long-range ordered Type I vertices, decimation introduces defects that locally disrupt this order and give rise to a heterogeneous network of frustrated configurations. This is reflected in the increase in vertex entropy and the growing population of higher-energy vertex types, indicating that the system explores a broader set of accessible microstates as the degree of decimation increases.

The dynamical measurements provide direct evidence for this crossover. At moderate decimation, the relaxation dynamics are well described by an Arrhenius form, indicating that spin reversals occur through thermally activated processes with a characteristic energy barrier. In this regime, the four-point susceptibility remains small and the Edwards--Anderson order parameter remains close to unity, consistent with weak dynamical heterogeneity and largely independent spin dynamics. In contrast, at higher decimation, the system exhibits a pronounced enhancement of the four-point susceptibility and a reduction of temporal correlations, signaling the emergence of dynamical heterogeneity. The corresponding relaxation times deviate from Arrhenius behavior and follow a Vogel--Fulcher form with a finite freezing temperature, indicating that relaxation proceeds through cooperative processes in an increasingly rugged energy landscape.

These observations place decimated artificial spin ice in close analogy with glass-forming systems, where the slowing down of dynamics is associated with the growth of correlated regions and the emergence of cooperative rearrangements. In particular, the increase in $\chi_{4}$ and the crossover to Vogel--Fulcher dynamics are hallmarks of glass formation, reflecting a transition from local to collective relaxation mechanisms. Importantly, in contrast to conventional spin glasses, where disorder is intrinsic and difficult to control, artificial spin ice enables disorder to be introduced in a systematic and tunable manner. This allows direct access to the relationship between defect density, frustration, and glassy dynamics.

More broadly, our results highlight the role of network connectivity and defect-induced interactions in governing the dynamical behavior of frustrated systems. Random decimation modifies not only the local coordination of spins but also the global connectivity of the interaction network, leading to a transition from a regime dominated by independent excitations to one characterized by cooperative dynamics and dynamical arrest. This perspective suggests that glass formation in frustrated magnetic systems can be understood in terms of the emergence of a correlated defect network that mediates collective relaxation processes.

Artificial spin ice therefore, provides a unique experimental platform for studying glassy dynamics with direct real-space access to both structure and dynamics. The ability to image individual spins, tune disorder, and probe dynamics across temperature scales offers opportunities to explore fundamental questions in nonequilibrium statistical physics, including the nature of dynamical heterogeneity, the role of defects in glass formation, and the interplay between frustration and disorder. Future work could extend these studies to different lattice geometries, interaction strengths, and disorder protocols, enabling systematic exploration of glass formation in engineered magnetic systems.

\section*{Acknowledgments}
The authors thank C.~Nisoli for fruitful discussions. This project was supported by the National Science Foundation Grant No.~2400155. Part of this work was carried out on the SIM beamline of the Swiss Light Source. F. C. is an employee of Planckian.

\paragraph*{Funding:}
 A.~F., D.~C. and S.~T received funding from the National Science Foundation Grant No.~2400155.
 
\paragraph*{Author contributions:}
A.F. designed the project and planned the experiments accordingly. K.H. fabricated the samples. D.C. performed the synchrotron experiments and analyzed the data. R.M., D.C. and A.F. analyzed the data. F.C. supported theoretical aspects of the project and analyzed the data. A.F., R.M. and D.C. wrote the manuscript with input from all co-authors.

\paragraph*{Competing interests:}
There are no competing interests to declare.
\paragraph*{Data and materials availability:}
Data relevant to this manuscript can be obtained from the authors upon request.


\subsection*{Materials and Methods}
\textbf{Sample fabrication.}
We use lift-off-assisted e-beam lithography to fabricate artificial square ice patterns consisting of Ising-type Ni$_{80\%}$Fe$_{20\%}$ (permalloy) nanomagnets with lengths of 400~nm and widths of 100~nm arranged on the sites of a background square lattice with lattice parameter 550~nm. The thickness $d$ of the nanomagnets was chosen to be 2.7~nm, which ensures thermally-activated moment reorientations to start emerging at the timescale of XMCD imaging (10 seconds per image) between 200 and 240~K~\cite{Farhan2019,Saccone2022}.

\textbf{X-ray PEEM.}
Measurements were performed at the photoemission electron microscope PEEM endstation at the SIM beamline of the Swiss Light Source. Magnetic images were recorded by taking advantage of XMCD at the Fe L$_3$ edge~\cite{Stohr1993}. The XMCD contrast is a measure of the projection of the magnetization on the x-ray polarization vector so that nanomagnets with a magnetization parallel or antiparallel to the x-ray polarization appear either black or white. Nanomagnets with moments having $\pm45^{\circ}$ and $\pm135^{\circ}$ angles with respect to the incoming x-rays appear dark and bright, respectively.

\textbf{Edwards--Anderson parameter and four-point susceptibility.}
To quantify the degree of temporal freezing and the emergence of dynamical heterogeneity in the decimated artificial spin ice, we analyze time-resolved XMCD image sequences recorded at fixed temperature. Each sequence consists of magnetic configurations acquired at time intervals of $\Delta t = 10$ s. From each XMCD image, the magnetic state of every nanomagnet is mapped onto an Ising variable $s_i(t)=\pm1$, corresponding to the two possible orientations of the magnetic moment along the long axis of the nanomagnet.

The Edwards--Anderson order parameter $q_{EA}$ was used to characterize the persistence of magnetic configurations over time. For each nanomagnet $i$, we first determine the time-averaged moment

\begin{equation}
\langle s_i \rangle_t = \frac{1}{N_t}\sum_{t=1}^{N_t} s_i(t),
\end{equation}

where $N_t$ is the number of frames in the time sequence. The Edwards--Anderson order parameter is then defined as

\begin{equation}
q_{EA}=\frac{1}{N}\sum_{i=1}^{N} \langle s_i \rangle_t^2 ,
\end{equation}

where $N$ is the total number of nanomagnets in the observed field of view. A value $q_{EA}\approx1$ indicates that most spins remain frozen in a fixed orientation throughout the measurement, whereas lower values correspond to increased fluctuations and a reduced memory of the initial configuration. The Edwards--Anderson parameter is a standard measure of freezing in spin glasses and disordered magnetic systems~\cite{Edwards1975,Binder1986}.

To quantify cooperative dynamics, we calculate the two-time autocorrelation function

\begin{equation}
C(t,\tau)=\frac{1}{N}\sum_{i=1}^{N} s_i(t)s_i(t+\tau),
\end{equation}

where $\tau$ is the lag time between two frames. The average autocorrelation is obtained by averaging over all possible initial times $t$,

\begin{equation}
\overline{C}(\tau)=\frac{1}{N_t-\tau}\sum_{t=1}^{N_t-\tau} C(t,\tau).
\end{equation}

The four-point susceptibility $\chi_4$ is then extracted from the temporal fluctuations of this two-time correlation function according to

\begin{equation}
\chi_4(\tau)=N\left[\left\langle C(t,\tau)^2\right\rangle_t-\left\langle C(t,\tau)\right\rangle_t^2\right],
\end{equation}

where the averages are again taken over all possible initial times $t$. Thus, $\chi_4$ measures the variance of the two-time autocorrelation and provides a measure of how strongly the dynamics are spatially correlated. Small values of $\chi_4$ correspond to nearly independent spin flips, whereas large values indicate that groups of spins rearrange cooperatively. The maximum value of $\chi_4(\tau)$ was used as a measure of the degree of dynamical heterogeneity at a given temperature. The corresponding lag time $\tau_{\mathrm{peak}}$ defines the characteristic timescale on which cooperative rearrangements occur. Such four-point susceptibilities are widely used in studies of structural and spin glasses to quantify growing dynamical correlations~\cite{Berthier2005,Berthier2011}. For each temperature and decimation level, $q_{EA}$, $\chi_4(\tau)$, and $\tau_{\mathrm{peak}}$ were extracted independently from up to 80 XMCD image sequences. The values shown in Figs.~4 and 5 correspond to the mean over all sequences, while the error bars represent the corresponding standard deviations.

\textbf{Entropy extraction from PEEM/XMCD moment configurations.}
To quantify the configurational disorder introduced by random decimation, we extracted entropies directly from the frozen magnetic microstates imaged after thermal annealing. The analysis was performed separately for local environments containing four, three, and two interacting nanomagnets, which we denote by $S_4$, $S_3$, and $S_2$, respectively. These local environments naturally emerge in the decimated square artificial spin ice because random removal of islands lowers the coordination of vertices surrounding a vacancy.

For each imaged configuration, every local motif of coordination number $z=4,3,2$ was identified from the binary spin map obtained from XMCD contrast. Each motif was assigned to one of its possible Ising microstates. For a motif with coordination $z$, the probability of observing microstate $\alpha$ was estimated from its measured frequency,
\begin{equation}
P_z(\alpha)=\frac{N_z(\alpha)}{\sum_{\alpha}N_z(\alpha)},
\end{equation}
where $N_z(\alpha)$ is the number of occurrences of state $\alpha$ among all motifs of coordination $z$ in a given image ensemble. From these probabilities, we computed the Shannon entropy associated with that coordination class,
\begin{equation}
S_z=-\frac{1}{z}\sum_{\alpha}P_z(\alpha)\log_2 P_z(\alpha),
\end{equation}
with $z=2,3,4$. The factor $1/z$ normalizes the entropy to units of bits per spin, allowing direct comparison between motifs of different coordination.

The total entropy per spin, plotted as $S$ in Fig.~\ref{fig:3}, was then obtained by combining the contributions from the different local coordinations according to their abundance in the decimated lattice,
\begin{equation}
S=\sum_{z=2}^{4} w_z S_z,
\end{equation}
where $w_z$ is the fraction of spins belonging to motifs of coordination $z$ in the corresponding decimated array. In this way, $S$ provides a global measure of configurational disorder, while $S_4$, $S_3$, and $S_2$ separately track the entropy associated with intact four-island vertices and decimation-induced reduced-coordination motifs.

For each decimation percentage, the entropy values were averaged over repeated experimentally imaged configurations obtained after identical annealing protocols. The error bars shown in Fig.~\ref{fig:3} correspond to the standard deviation over these repetitions. Entropies are reported in bits per spin.


\begin{thebibliography}{}
\bibitem{Binder1986} K. Binder, and A.P. Young, Spin glasses: Experimental facts, theoretical concepts, and open questions, Reviews of Modern Physics {\bf 58}, 801-976 (1986). 
\bibitem{Debenedetti2001} P. G. Debenedetti, and Frank H. Stillinger, Supercooled liquids and the glass transition, Nature {\bf 410}, 259--267 (2001).
\bibitem{Berthier2011} L. Berthier, and G. Biroli, Theoretical perspective on the glass transition and amorphous materials, Reviews of Modern Physics {\bf 83}, 587 (2011).
\bibitem{Ediger2000} M. D. Ediger, Spatially Heterogeneous Dynamics in Supercooled Liquids, Annual Review of Physical Chemistry {\bf 51}, 99-128 (2000).
\bibitem{Angell1995} C. A. Angell, Formation of Glasses from Liquids and Biopolymers, Science {\bf 267}, 1924-1935 (1995).
\bibitem{Cavagna2009} A. Cavagna, Supercooled liquids for pedestrians, Physics Reports {\bf 476}, 51-124 (2009).
\bibitem{Saccone20232}
M. Saccone and F. Caravelli,
Complex field reversal dynamics in nanomagnetic systems,
\textit{Proceedings of the Royal Society A: Mathematical, Physical and Engineering Sciences} \textbf{479}, 2277 (2023).
\bibitem{Halln2022}
J. N. Hall\'en, S. A. Grigera, D. A. Tennant, C. Castelnovo, and R. Moessner,
Dynamical fractal and anomalous noise in a clean magnetic crystal,
\textit{Science} \textbf{378}, 1218--1221 (2022).
\bibitem{Harris2020} S. T. Bramwell, and M. J. Harris, The history of spin ice, J. Phys.: Condens. Matter {\bf 32}, 374010 (2020).
\bibitem{Skjaervo2020} S. H. Skjærvø, C. H. Marrows, R. L. Stamps, and L. J. Heyderman, Advances in artificial spin ice, Nature Reviews Physics {\bf 2}, 13--28 (2020).
\bibitem{Farhan2017} A. Farhan, C. F. Petersen, S. Dhuey, L. Anghinolfi, Q. H. Qin, M. Saccone, S. Velten, C. Wuth, S. Gliga, P. Mellado, M. J. Alava, A. Scholl, and S. van Dijken,  Nanoscale control of competing interactions and geometrical frustration in a dipolar trident lattice, Nature Communications {\bf 8}, 995 (2017).
\bibitem{Farhan2019} A. Farhan, M. Saccone, C. F. Petersen, S. Dhuey, R. V. Chopdekar, Y.-L. Huang, N. Kent, Z. Chen, M. J. Alava, T. Lippert, A. Scholl, S. van Dijken, Emergent magnetic monopole dynamics in macroscopically degenerate artificial spin ice, Science Advances {\bf 5}, eaav638 (2019).
\bibitem{Gilbert2014} I. Gilbert, G.-W. Chern, S. Zhang, L. O’Brien, B. Fore, C. Nisoli, and P. Schiffer, Emergent ice rule and magnetic charge screening from vertex frustration in artificial spin ice, Nature Physics {\bf 10}, 670–675 (2014).
\bibitem{Gilbert2015} I. Gilbert, Y. Lao, I. Carrasquillo, L. O’Brien, J. D. Watts, M. Manno, C. Leighton, A. Scholl, C. Nisoli, and P. Schiffer, Emergent reduced dimensionality by vertex frustration in artificial spin ice, Nature Physics {\bf 12}, 162--165 (2016).
\bibitem{Lao2018} Y. Lao, F. Caravelli, M. Sheikh, J. Sklenar, D. Gardeazabal, J. D. Watts, A. M. Albrecht, A. Scholl, K. Dahmen, C. Nisoli, P. Schiffer, Classical topological order in the kinetics of artificial spin ice, Nature Physics {\bf 14}, 723--727 (2018).
\bibitem{Leo2018} N. Leo, S. Holenstein, D. Schildknecht, O. Sendetskyi, H. Luetkens, P. M. Derlet, V. Scagnoli, D. Lançon, J. R. L. Mardegan, T. Prokscha, A. Suter, Z. Salman, S. Lee, L. J. Heyderman, Collective magnetism in an artificial 2D XY spin system, Nature Communications {\bf 9}, 2850 (2018).
\bibitem{Louis2018} D. Louis, D. Lacour, M. Hehn, V. Lomakin, T. Hauet, and F. Montaigne, A tunable magnetic metamaterial based on the dipolar four-state Potts model, Nature Materials {\bf 17}, 1076–1080 (2018).
\bibitem{Perrin2016} Y. Perrin, B. Canals, and N. Rougemaille, Coulomb phase and magnetic monopoles in artificial square ice, Nature {\bf 540}, 410–413 (2016).
\bibitem{Ladak2010} S. Ladak, D. E. Read, G. K. Perkins, L. F. Cohen, and W. R. Branford, Direct observation of magnetic monopole defects in an artificial spin-ice system, Nature Physics {\bf 6}, 359–363 (2010).
\bibitem{Saccone2020} M. Saccone, K. Hofhuis, D. Bracher, A. Kleibert, S. van Dijken,  and  A. Farhan, Elevated effective dimension in tree-like nanomagnetic Cayley structures, Nanoscale {\bf 12}, 189-194 (2020).
\bibitem{Saccone2022} M. Saccone, F. Caravelli, K. Hofhuis, S. Parchenko, Y. A. Birkhölzer, S. Dhuey, A. Kleibert, S. van Dijken, C. Nisoli, and A. Farhan, Direct observation of a dynamical glass transition in a nanomagnetic artificial Hopfield network, Nature Physics {\bf 18}, 517–521 (2022).
\bibitem{Sen2015} A. Sen, and R. Moessner, Topological Spin Glass in Diluted Spin Ice, Physical Review Letters {\bf 114}, 247207 (2015).
\bibitem{Sala2014} G. Sala, M. J. Gutmann, D. Prabhakaran, D. Pomaranski, C. Mitchelitis, J. B. Kycia, D. G. Porter, C. Castelnovo, and J. P. Goff, Vacancy defects and monopole dynamics in oxygen-deficient pyrochlores, Nature Materials {\bf 13}, 488–493 (2014).
\bibitem{Stohr1993} J. St\"{o}hr, Y. Wu, B. D. Hermeister, M. G. Samant, G. R. Harp, S. Koranda, D. Dunham, and B. P. Tonner, Element-specific magnetic microscopy with circularly polarized x-rays, Science {\bf 259},  658-661 (1993).
\bibitem{Crater2024} D. Crater, A. C. Lätti, M. Saccone, K. Hofhuis, D. Miertschin, B. Regmi, B. Achinuq, A. Kleibert, C. F. Petersen, and A. Farhan, Ice-rule driven vertex frustration in a stretched pentagonal spin ice, Physical Review Research {\bf 6}, L042064 (2024).
\bibitem{Morrison2013} M. J. Morrison, T. R. Nelson, and C. Nisoli, Unhappy vertices in artificial spin ice: new degeneracies from vertex frustration, New Journal of Physics {\bf 15}, 045009 (2013).
\bibitem{Saccone2023} M. Saccone, F. Caravelli, K. Hofhuis, S. Dhuey, A. Scholl, C. Nisoli, and A. Farhan, Real-space observation of ergodicity transitions in artificial spin ice, Nature Communications {\bf 14}, 5674 (2023).
\bibitem{Moller2009} G. M\"oller and R. Moessner, Magnetic multipole analysis of kagome and artificial spin-ice dipolar arrays, Physical Review B {\bf 80}, 140409(R) (2009).
\bibitem{Kagome2017} A. Farhan, P. M. Derlet, L. Anghinolfi, A. Kleibert, and L. J. Heyderman, Magnetic charge and moment dynamics in artificial kagome spin ice, Physical Review B {\bf 96}, 064409 (2017).
\bibitem{Lammert2010} P. E. Lammert, X. Ke, J. Li, C. Nisoli, D. M. Garand, V. H. Crespi, and P. Schiffer, Direct entropy determination and application to artificial spin ice, Nature Physics {\bf 6}, 786--789 (2010).
\bibitem{Moller2006} G. M\"oller and R. Moessner, Artificial Square Ice and Related Dipolar Nanoarrays, Physical Review Letters {\bf 96}, 237202 (2006).
\bibitem{Alan2013} A. Farhan, P. M. Derlet, A. Kleibert, A. Balan, R. V. Chopdekar, M. Wyss, J. Perron, A. Scholl, F. Nolting, and L. J. Heyderman, Direct observation of thermal relaxation in artificial spin ice, Physical Review Letters {\bf 111}, 057204 (2013).
\bibitem{Budrikis2010} Z. Budrikis, P. Politi, and R. L. Stamps, Vertex Dynamics in Finite Two-Dimensional Square Spin Ices, Physical Review Letters {\bf 105}, 017201 (2010).
\bibitem{Edwards1975} S. F. Edwards and P. W. Anderson, Theory of Spin Glasses, J. Phys. F: Met. Phys. {\bf 5}, 965 (1975).
\bibitem{Berthier2005} L. Berthier, G. Biroli, J.-P. Bouchaud, L. Cipelletti, D. El Masri, D. L'H\^ote, F. Ladieu, and M. Pierno, Direct Experimental Evidence of a Growing Length Scale Accompanying the Glass Transition, Science {\bf 310}, 1797-1800 (2005).
\bibitem{Morley2017} S. A. Morley, D. Alba Venero, J. M. Porro, S. T. Riley, A. Stein, P. Steadman, R. L. Stamps, S. Langridge, and C. H. Marrows, Vogel-Fulcher-Tammann freezing of a thermally fluctuating artificial spin ice probed by x-ray photon correlation spectroscopy, Physical Review B {\bf 95}, 104422 (2017).

\end{thebibliography}
\end{document}